\documentclass[aps,prl,twocolumn,amsmath,amssymb,nofootinbib,superscriptaddress]{revtex4-1}
\usepackage{graphicx}
\usepackage{epstopdf}

\usepackage{hyperref}
\usepackage{cleveref}

\newcommand{\bra}[1]{\langle#1|}
\newcommand{\ket}[1]{|#1\rangle}

\crefname{equation}{Eq.}{Eqs.}
\Crefname{equation}{Equation}{Equations}
\crefname{figure}{Fig.}{Figs.}
\Crefname{figure}{Figure}{Figures}
\crefname{section}{Sec.}{Secs.}
\crefname{subsection}{Subsec.}{Subsecs.}
\Crefname{section}{Section}{Sections}
\crefname{appendix}{Appendix}{Appendices}
\Crefname{appendix}{Appendix}{Appendices}

\DeclareMathOperator{\circleright}{\rotatebox[]{-90}{$\circlearrowright$}}

\DeclareMathOperator{\circleleft}{  \rotatebox[] {90}{$\circlearrowleft$}}

\newcommand{\IL}{{\rm IL}}
\newcommand{\IS}{{\rm IS}}
\newcommand{\R}{{\rm R}}
\newcommand{\eee}{\textsf{eee}}
\newcommand{\eoo}{\textsf{eoo}}
\newcommand{\oeo}{\textsf{oeo}}
\newcommand{\ooe}{\textsf{ooe}}

\begin{document}
\title[Article Title]{Passive microwave circulation on a superconducting chip}
\author{Arkady Fedorov}
\affiliation{Analog Quantum Circuits Pty.\ Ltd., Brisbane, Australia}
\affiliation{School of Mathematics and Physics, University of Queensland, Brisbane, QLD 4072, Australia}
\author{N. Pradeep Kumar}
\affiliation{Analog Quantum Circuits Pty.\ Ltd., Brisbane, Australia}
\affiliation{School of Mathematics and Physics, University of Queensland, Brisbane, QLD 4072, Australia}
\author{Dat Thanh Le} 
\affiliation{Analog Quantum Circuits Pty.\ Ltd., Brisbane, Australia}
\affiliation{School of Mathematics and Physics, University of Queensland, Brisbane, QLD 4072, Australia}
\author{\mbox{Rohit Navarathna}}
\affiliation{Analog Quantum Circuits Pty.\ Ltd., Brisbane, Australia}
\affiliation{School of Mathematics and Physics, University of Queensland, Brisbane, QLD 4072, Australia}
\author{Prasanna Pakkiam}
\affiliation{Analog Quantum Circuits Pty.\ Ltd., Brisbane, Australia}
\affiliation{School of Mathematics and Physics, University of Queensland, Brisbane, QLD 4072, Australia}
\author{Thomas M. Stace} 
\affiliation{Analog Quantum Circuits Pty.\ Ltd., Brisbane, Australia}
\affiliation{School of Mathematics and Physics, University of Queensland, Brisbane, QLD 4072, Australia}
\begin{abstract}

Building large-scale superconducting quantum circuits will require miniaturisation and integration of supporting devices including microwave circulators, which are currently bulky, stand-alone components.  Here we report the realisation of a passive on-chip circulator which is made from a loop consisting of three tunnel-coupled superconducting islands, with DC-only control fields.  We observe the effect of quasiparticle tunnelling, and we dynamically classify the system into different quasiparticle sectors.  When tuned for circulation, the device exhibits strongly
non-reciprocal 3-port scattering, with average on-resonance insertion loss of 2 dB, isolation of $14$ dB, power reflectance of $-11$ dB, and a bandwidth of 200 MHz.  \end{abstract}

\maketitle


Circulators are non-reciprocal multi-port devices used to route electromagnetic signals \cite{Kord18}, and are ubiquitous in cryogenic microwave circuits \cite{Pozar11,Gu17} for isolating  a system-under-test from thermal noise \cite{metelmannNonreciprocalPhotonTransmission2015,Ruesink16}.  Conventional ferrite microwave circulators are centimeter-scale, magnetised units which are not amenable to microfabrication and integration on chip, and thus present a constraint on the development of large-scale solid-state quantum processors.  
Various approaches to miniaturising non-reciprocal signal routing have been proposed, including actively driven systems which require additional radio-frequency or microwave control fields \cite{chapmanWidelyTunableOnChip2017, kamalNoiselessNonreciprocityParametric2011, kamalMinimalModelsNonreciprocal2017, estepMagneticfreeNonreciprocityIsolation2014, sliwaReconfigurableJosephsonCirculator2015, lecocqNonreciprocalMicrowaveSignal2017, fangGeneralizedNonreciprocityOptomechanical2017, metelmannNonreciprocalPhotonTransmission2015, petersonStrongNonreciprocityModulated2019, kerckhoffOnChipSuperconductingMicrowave2015, roushanChiralGroundstateCurrents2017, rosenthalBreakingLorentzReciprocity2017} and quantum-Hall based devices which require large magnetic fields \cite{PhysRevLett.93.126804,violaHallEffectGyrators2014, mahoneyOnChipMicrowaveQuantum2017}.  

\begin{figure}[t]
    \centering
    \includegraphics[]{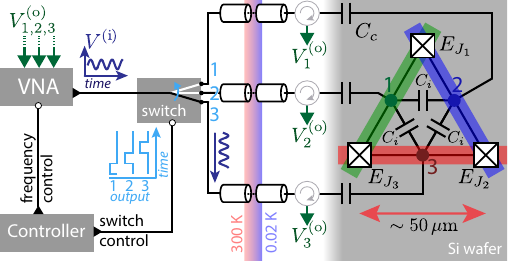}
    \caption{Circuit diagram for the experimental system.  The core device is a ring of three overlapping aluminium islands [green, blue and red boxes], each  $\sim50\,\mu{\rm m}$ long, which are mutually tunnel coupled by junctions with Josephson energies $E_{J_{1,2,3}}$.  The islands couple capacitively to each other,  $C_{i}$, to external waveguides and DC-biases, $C_c$, and to ground,  $C_g$ (not shown).  A controller sets the drive  signal $V^{(\rm i)}$ from a vector network analyser (VNA), and a fast microwave switch that directs the signal to one of the three waveguide inputs.  The scattered signals $V^{(\rm o)}_{1,2,3}$ are directed back to the VNA.}
    \label{fig:circuit}
\end{figure}

Here, we report the experimental observation of microwave circulation in a passive, on-chip superconducting device first proposed by \citet{kochTimereversalsymmetryBreakingCircuitQEDbased2010}, which consists  of  three superconducting,  tunnel-coupled, aluminium  islands arranged in a ring topology. 
This microfabricated, on-chip device is predicted to exhibit high-performance microwave circulation  without large magnetic  or dynamical control fields \cite{mullerPassiveOnChipSuperconducting2018,leOperatingPassiveOnchip2021}, making it a promising candidate for miniaturising and integrating microwave circulators with other superconducting devices on the same wafer.  
The experimental system is represented in \cref{fig:circuit}, including the three  aluminium islands, indicated by the green, blue and red boxes, which are deposited on a silicon wafer.  The islands are capacitively coupled to one another, to ground, and to the external waveguides through which the system is driven by input signals, $V^{\rm(i)}_{1,2,3}$.  The scattered output, $V^{\rm(o)}_{1,2,3}$, is used to determine the  scattering matrix amplitudes, $S_{ab}=V^{\rm(o)}_b/V^{\rm(i)}_a$, with $a,b\in\{1,2,3\}$.  
The system response depends on the driving frequency, the external flux bias, and the DC charge bias applied to the superconducting islands.

Our previous modelling of this system, based on the `SLH' formalism \cite{Combesdoi:10.1080/23746149.2017.1343097}, quantises  the nodal flux and charge at each island, $\hat \phi_{1,2,3}$ and $ \hat n_{1,2,3}$ respectively, and constructs a Hamiltonian, $H_{\rm ring}$, which is capacitively coupled to three waveguides \cite{mullerPassiveOnChipSuperconducting2018,leOperatingPassiveOnchip2021}.    
Recently, we  validated our theoretical model against experimental results in a different device design, establishing good quantitative agreement between theory and experiment \cite{PhysRevLett.130.037001}.  We adopt the same theoretical model to analyse the experimental results reported here. 
Briefly, the  Hamiltonian describing the ring is parameterised by the three Josephson tunnel-junction energies, $E_{J_{1,2,3}}$, and the lumped-element capacitance matrix, which includes capacitances between the metallic islands, $C_{i}$, the waveguides and the islands, $C_{c}$, and the island capacitances to ground, $C_{g}$.

For ideal circulation, the ring Hamiltonian should be symmetric under cyclic permutations of the node labels, $1\rightarrow2\rightarrow3\rightarrow1$, requiring the three islands to be electrically symmetric.  This requires the junction energies to be identical, and the system's capacitance matrix to be symmetric. 
In practice, the electrical symmetry is broken by fabrication variations, for example leading to a spread in the actual $E_J$'s or  $C_i$'s.  Our earlier modelling predicted good circulation when the spread in junction energies was within 1\% of the design values 
\cite{mullerPassiveOnChipSuperconducting2018, leOperatingPassiveOnchip2021}.

The ring device is voltage-tunable, so is sensitive to charge fluctuations.  One of the key empirical observations in \citet{PhysRevLett.130.037001} was the presence of discrete charge fluctuations that were well-described by a \mbox{$K=4$ -state} Hidden markov model with state lifetimes $\sim200\,\mu$s.  These hidden states were hypothesised to arise from quasiparticle tunnelling between the islands, which  generates four distinct quasiparticle sectors, labelled as $\eee, \eoo, \oeo$, and $\ooe$ \cite{leOperatingPassiveOnchip2021}.  The reference configuration, $\eee$ consists of an even quasi-particle parity [$\textsf{e}$]  on each island; a quasiparticle-tunnelling event changes this to odd parity [$\textsf{o}$] on two of the islands.  

To quantify the circulation performance of the device,  we define  the average clockwise and anticlockwise circulation fidelities, and the average reflection respectively as 
\begin{subequations}
\begin{eqnarray}
{\mathcal F}_{\,\circleright}&=&(|S_{12}|+|S_{23}|+|S_{31}|)/3,\\
{\mathcal F}_{\,\circleleft}&=&(|S_{13}|+|S_{32}|+|S_{21}|)/3,\\
{\mathcal R}&=&(|S_{11}|+|S_{22}|+|S_{33}|)/3.
\end{eqnarray}
\label{eqn:sums}
\end{subequations}
An ideal clockwise circulator will have ${\mathcal F}_{\,\circleright}=1$.

A conservative scattering element is described by a unitary matrix  $S=e^{i G}$, with Hermitian generator $G$.  In addition, if the scattering is time-reversal symmetric, then $G=G^*=G^T$ and $S$ will be symmetric, $S=S^T$. 
It is straightforward to show that for a time-reversal-symmetric $3\times3$ scattering matrix, ${\mathcal F}\leq2/3$. It follows that a scattering element with ${\mathcal F}>2/3$ exhibits time-reversal-symmetry breaking suitable for non-reciprocal scattering.

\begin{figure}[t]
    \centering
    \includegraphics[]{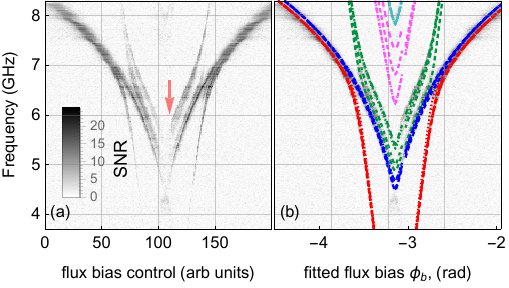}
    \caption{(a) Measured spectral response of the microwave voltage transmission from port $a=1$ to port $b=2$, $V_{12}$, as a function of flux bias control. 
    For each frequency, the raw output voltage data is scaled so that the off-resonant background data has zero mean and unit variance; the grey-scale therefore represents the output signal-to-noise ratio (SNR).  There is a  noticeable `glitch' indicated by the  arrow, with subtle  differences between the left and right halves of the spectrum.  (b) Fitted model spectrum superimposed on the measured spectrum, including 4 distinct quasiparticle sectors.  The fitted model includes a linear scale factor to convert the abscissa flux bias control parameter (an external control voltage) into a dimensionless bias flux, $\phi_b=2\pi\Phi_{b}/\Phi_0$.  Different colours correspond to modelled transition frequencies from the ground state to the first excited state [red], second [blue], third [green], fourth [pink] and fifth [aqua]; the four-fold multiplicity within the predicted spectrum arises from the different quasiparticle sectors.  To account for the glitch in panel (a), we allow for different charge bias configurations on the left and right halves of the modelled spectrum, consistent with a local charge shift that occurred roughly halfway through the data collection.}
    \label{fig:spectrum}
\end{figure}

\begin{figure}[]
    \centering
    \includegraphics[]{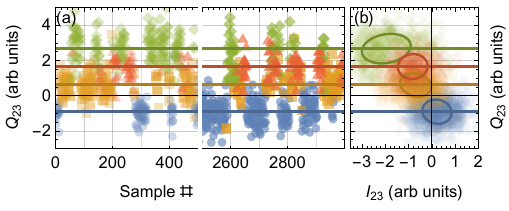}
    \caption{(a) Leading and trailing subsets of the time series data of the quadrature voltage signal, $Q_{23}$, transferred from port $a=2$ to port $b=3$ for 3000 measurement samples taken with sample measurement time $\tau_s=300\,\mu$s.  The samples are classified into one of $K=4$ sub-populations, indicated by point colour and shape, using a $K$-means classifier, with population means indicated by horizontal lines.  We identify the sub-populations with different quasiparticle sectors.   (b) Projection of the 3000 complex voltage amplitudes $V_{23}=I_{23}+i Q_{23}$ classified by sector, together with 1$\sigma$ covariance ellipses centered on each sector mean.}
    \label{fig:time_series}
\end{figure}

The results we report here are based on a device that was designed and computationally optimised to have highly symmetric capacitances and Josephson energies. Fabrication was undertaken as in \citet{PhysRevLett.130.037001}, using electron-beam lithography to pattern the design on a bi-layer resist stack.  Standard double-angle evaporation was then used to deposit two layers of aluminium on a high-resistivity silicon substrate, with a single oxidation step between the two aluminium deposition stages to grow the Josephson tunnel barriers in the device.  After evaporation, the chip was cleaved and bonded on a holder suitable for  cryogenic measurements in a dilution refrigerator operating at a base temperature of \mbox{20 mK}.

To characterise the system, we first measure the spectral response.  \Cref{fig:spectrum}(a) shows the signal-to-noise ratio (SNR) of the voltage transmission from port 1 to port 2, $V_{12}$, as a function of drive frequency, and a flux-bias control parameter.  The characteristic $\textsf{Y}$-shape of the spectrum has been predicted \cite{mullerPassiveOnChipSuperconducting2018, leOperatingPassiveOnchip2021} and seen experimentally \cite{PhysRevLett.130.037001} in our earlier work, as has the multiplets of lines associated to quasiparticle sectors.  Ideally, the spectrum should be symmetric under inversion of the flux bias (so that the left and right halves of the spectrum should be reflected), but this symmetry is broken in the data shown: we see a single `glitch' indicated by the arrow, with subtle but distinct variations in the multiplet structure on the left and right of the plot.

\begin{figure}[!t]
    \centering
    \includegraphics[]{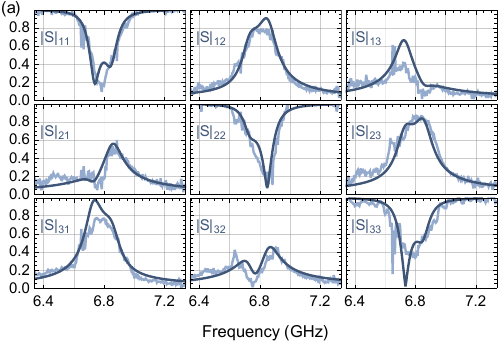}
    \includegraphics[]{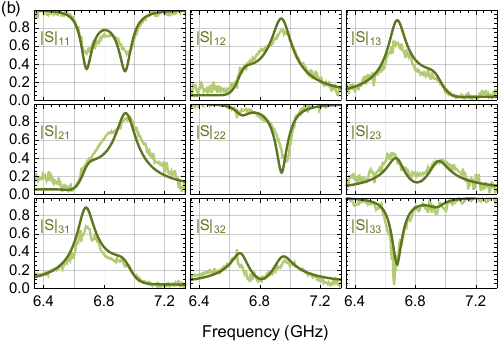}
    \includegraphics[]{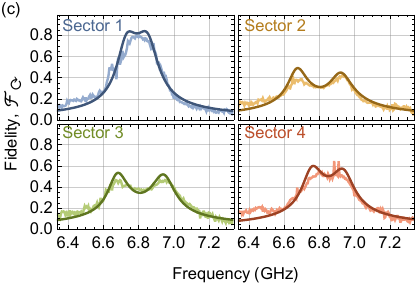}
    \caption{(a) Scattering matrix elements, $|S|_{ab}$, for time-series data classified as sector 1, which shows strong asymmetry in the scattering matrix, $S\neq S^T$, near 6.8 GHz. (b) Scattering matrix elements for sector 3, which is quite symmetric for all frequencies,  $S\approx S^T$.  Experimental data [light] shows good agreement with model calculations [dark].  Scattering matrix spectra for sectors 2 and 4 look qualitatively similar to (b).  (c) The clockwise circulation fidelity, 
   { ${\mathcal F}_{\,{\circleright} }$}, 
    of scattering matrices in each sector, showing experimental data and model predictions. 
    }
    \label{fig:smatrix}
\end{figure}

\Cref{fig:spectrum}(b) shows the spectrum predicted by our theoretical model, which fitted to, and superimposed upon, the same SNR data, with the horizontal axis rescaled into units of dimensionless flux bias $\phi_b=2\pi\Phi_b/\Phi_0$, where \mbox{$\Phi_0=h/(2e)$}.  It shows the predicted transition frequencies from the ground state to different excited states (colours).  
The  model fitting gives an on-site capacitive energy \mbox{$E_{C_{\Sigma}}/h=(2e)^2/(h C_{\Sigma})=2.97$ GHz}, corresponding to a total island capacitance of \mbox{$C_\Sigma=C_{g}+C_c+3C_i=52$ fF} which is consistent with the designed capacitances (\mbox{$C_g=3.5$ fF},  \mbox{$C_c=25$ fF}, and \mbox{$C_i=8$ fF})  and Josephson energies \mbox{$E_{J_{1,2,3}}/h=11.8, 11.8$} and \mbox{$12.06$ GHz}.

The Josephson energy is inversely related to the room-temperature junction resistance,  $E_J\propto 1/R_J$ \cite{Ambegaokar1963}.  We measured $R_{J_{1,2,3}}=11.37, 11.35$ and \mbox{11.16 k$\Omega$} respectively for this device. 
The spread of 1.9\% in $R_J$'s is consistent with the 2.2\% spread of $E_J$'s found above.

The fact that the theoretical spectrum shown in \cref{fig:spectrum}(b)  obscures the salient features in the SNR data demonstrates that the model explains the spectral data well.  Within each transition band, there are four distinct curves, corresponding to the different quasiparticle sectors.  For some transitions (e.g.\ the red and blue lines), these are nearly degenerate, and in others they are more distinct, but in all cases the underlying multiplets in the data are explained by the theoretical model.

Next, we measure the full $3\times3$ complex scattering matrix.  We do this using a fast microwave switch to sequentially direct the drive to each of the three input ports for 100 $\mu$s, and measure the three output ports with a vector network analyser, as represented in \cref{fig:circuit}; each sample of $S$ takes $\tau_s=300\,\mu$s.  We do this for different drive frequencies, bias voltages, and bias currents, to yield a large set of time-series data. 

\Cref{fig:time_series}(a) shows the imaginary voltage amplitude, $Q_{23}$ (where $V_{ab}=I_{ab}+i Q_{ab}$ for transmission from port $a$ to port $b$), for a time-series spanning 3000 samples, at a fixed drive frequency, bias voltage, and flux.  We see characteristic jumps between discrete output voltage states   \cite{PhysRevLett.130.037001}.  
We use a $K$-means classifier to separate the discrete voltage states in the time-series into $K=4$ statistically distinct Gaussian sub-populations, which are depicted in \cref{fig:time_series}(a) with different colours. Each sub-population is characterised by a mean and covariance \cite{schreiber2018pomegranate}, and a characteristic dwell time ranging from 10 to 22 samples (i.e.\ 3 to 6.6 ms).  We also show a projection of the sub-populations into the complex $V_{23}$ plane in \cref{fig:time_series}(b), together with the projected 1$\sigma$ sample-covariance ellipses for each.  (Note that the classifier simultaneously analyses all 9 complex-valued voltages $V_{ab}$, so that the statistical distance between the populations in this time-series is substantially larger than the projection shown in \cref{fig:time_series}(b).)  We attribute these four sub-populations to the four quasiparticle sectors described previously \cite{leOperatingPassiveOnchip2021,PhysRevLett.130.037001}.

Next, we fix external bias voltages and fluxes at a  working point with high circulation fidelity, and  measure scattering data while scanning the  drive frequency.  This yields scattering matrix spectra for each of the four sectors, which are shown for sector 1 in \cref{fig:smatrix}(a), and for sector 3 in \cref{fig:smatrix}(b), (scattering spectra for sectors 2 and 4 are similar to sector 3).  The most important comparative feature of the frequency responses is that the measured scattering matrix for sector 1 is strongly asymmetric, $S\neq S^T$ (e.g.\ around \mbox{6.8 GHz}, $|S_{12}|>|S_{21}|$, $|S_{23}|>|S_{32}|$ and $|S_{31}|>|S_{13}|$), whereas the scattering matrix for sector 3 is approximately symmetric, $S\approx S^T$.  The scattering asymmetry in \cref{fig:smatrix}(a) shows that the device circulates  when it is in the state corresponding to sector 1, while \cref{fig:smatrix}(b) shows that it does not circulate in the other sectors, consistent with the quasiparticle analysis in \citet{leOperatingPassiveOnchip2021}.

\Cref{fig:smatrix} also shows model predictions (darker curves), using the same circuit parameters used to generate the spectra in \cref{fig:spectrum}(b), with a fitted coupling strength at $\kappa=119\, {\rm MHz}$ for each waveguide.  The only model parameter we vary between the theory curves in  \cref{fig:smatrix}(a) and \cref{fig:smatrix}(b) is the offset charge bias on two of the islands, consistent with a discrete change in charge state due to quasiparticles in these two sectors.  We see good agreement between the data and the model predictions, with resonance frequencies and strengths reasonably well matched in each sector.

\begin{figure}
    \centering
    \includegraphics[]{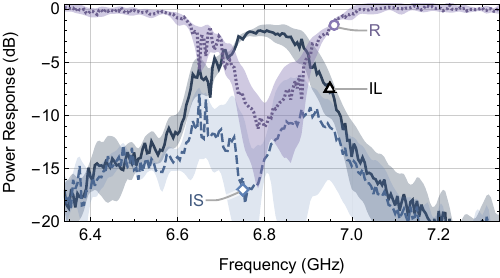}
    \caption{Average power-transfer matrix-elements for \mbox{sector 1} data,  showing the average insertion loss, 
    $\IL={\mathcal F}_{\,\circleright}^2$, isolation,  
    $\IS={\mathcal F}_{\,\circleleft}^2$ and power
     reflectance $\R={\mathcal R}^2$. 
     The shaded region around each quantity indicates its range, estimated from the smoothed maxima and minima over each of the terms in the corresponding sums in \cref{eqn:sums}.
    }
    \label{fig:circulation}
\end{figure}

We compare circulation in the four sectors using the circulation fidelity measure ${\mathcal F}_{\,\circleright}$, shown in \cref{fig:smatrix}(c).  We see high fidelity clockwise circulation in sector 1, reaching $\max {\mathcal F}_{\,\circleright}=0.8$, well above the ${\mathcal F}\leq 2/3$ bound for time-reversal symmetric devices, and with a bandwidth of \mbox{$200$ MHz} at full-width--half-maximum.   Conversely, the other sectors have maximum fidelities lower than the time-reversal-symmetric threshold, and do not show strong circulation.  In all sectors, the circulation fidelity inferred from the measurements is in good agreement with the model using  parameters obtained from the spectral fitting. The consistency between the modelling and experimental data indicate that the observed circulation is limited by the $\sim2\%$ variation in the junction energies, $E_{J_{1,2,3}}$.

We are able to tune the device operation by application of DC voltage and flux biases.  These enable us to dynamically reverse the direction of circulation, which we have observed in this device.  The results for anticlockwise circulation are essentially identical to those presented here.

Finally, we characterise the device performance as a (clockwise) circulator.   
From the scattering matrix and the quantities defined in \cref{eqn:sums}, we define  the average insertion loss \mbox{$\IL={\mathcal F}_{\,\circleright}^2$}, the average isolation \mbox{$\IS={\mathcal F}_{\,\circleleft}^2$}, and the average power reflectance  \mbox{$\R={\mathcal R}^2$}.  We plot these  in \cref {fig:circulation}, and see that at the \mbox{$6.8$ GHz} resonance frequency, \mbox{$\IL=2$ dB}, \mbox{$\IS=14$ dB} and \mbox{$\R=-11$ dB}.  

The device saturation power will roughly correspond to the arrival of one drive photon per excited-state lifetime, $\tau_e=(\kappa \,|\bra{e}\hat n_a\ket{g}|^2)^{-1}$, where we use the model to compute $|\bra{e}\hat n_a\ket{g}|^2=0.38$, based on fitted device parameters.  This estimate gives \mbox{$P_{\rm sat}\approx h f/ \tau_e=-127$ dBm}.  Consistent with this estimate, the full model simulation predicts the circulation fidelity will be halved relative to the zero power limit when \mbox{$P_{\rm in}=-127$ dBm}. 

To conclude, we have built a DC-controlled microwave circulator using a ring of tunnel-coupled superconducting islands driven through external waveguides. We observe discrete charge jumps that we classify as quasiparticle tunnelling events through the junctions.  
When the system is in the optimised quasiparticle sector we observe strong circulation in the microwave scattering matrix.  Our measured results are in good quantitative agreement with model predictions, which provides guidance on pathways to improve performance. 

\begin{acknowledgments}
This work was funded through a commercial research contract with Analog Quantum Circuits (AQC) Pty.\ Ltd.  TMS and AK each declare a financial interest in AQC. 
The authors acknowledge assistance from the Centre for Microscopy and Microanalysis at the University of Queensland, and the Australian National Fabrication Facility, ANFF-Q.
\end{acknowledgments}

\bibliography{paper}

\end{document}